\documentclass{IEEEtran}
\usepackage{graphicx}
\usepackage{amsmath}
\usepackage{multicol}
\usepackage{multirow}
\usepackage{stfloats}
\usepackage{booktabs}
\usepackage{mathrsfs}
\usepackage{enumerate}
\usepackage{amssymb}
\usepackage{algorithm}
\usepackage{algpseudocode}

\usepackage{array}
\usepackage{underscore}
\usepackage{url}

\usepackage{hyperref}
\hypersetup{hidelinks}
\usepackage{flushend}

\usepackage[square, comma, sort&compress, numbers]{natbib}
\usepackage{graphicx}
\usepackage{epstopdf}
\usepackage{caption}
\usepackage{diagbox}

\usepackage{caption}
\usepackage{subfigure}
\usepackage{color}

\usepackage[outerbars,color]{changebar}
\ifx\pdfoutput\undefined
\else\ifnum\pdfoutput>0
  \usepackage{pdfcolmk}
\fi\fi
\cbcolor{red}
\usepackage{fancyhdr}

\makeatletter
\newcommand{\Rmnum}[1]{\expandafter\@slowromancap\romannumeral #1@}
\makeatother

\usepackage{array}  \newcommand{\PreserveBackslash}[1]{\let\temp=\\#1\let\\=\temp}  \newcolumntype{C}[1]{>{\PreserveBackslash\centering}p{#1}}  \newcolumntype{R}[1]{>{\PreserveBackslash\raggedleft}p{#1}}  \newcolumntype{L}[1]{>{\PreserveBackslash\raggedright}p{#1}}

\ifCLASSINFOpdf
\else
\fi
\let\oldbibitem\bibitem
\def\bibitem{\vfill\oldbibitem}

\let\oldhref\href
\renewcommand{\href}[2]{\oldhref{#1}{\hbox{#2}}}

\begin{document}

\title{\begin{Huge}Earning Maximization with Quality of Charging Service Guarantee for IoT Devices\end{Huge}}

\author{Wen~Fang,
Qingqing~Zhang,
Mingqing~Liu,
Qingwen~Liu\IEEEauthorrefmark{1},
and~Pengfei~Xia	

	\thanks{W.~Fang, Q. Zhang, M. Liu, Q. Liu, and P. Xia are with College of Electronic and Information Engineering, Tongji University, Shanghai, People's Republic of China, (e-mail: wen.fang@tongji.edu.cn, anne@tongji.edu.cn, clare@tongji.edu.cn, qliu@tongji.edu.cn, and pengfei.xia@gmail.com).}%
    \thanks{* Corresponding author.}
}

\maketitle

\begin{abstract}
Resonant Beam Charging (RBC) is a promising Wireless Power Transfer (WPT) technology to provide long-range, high-power, mobile and safe wireless power for the Internet of Things (IoT) devices. The Point-to-Multipoint (PtMP) RBC system can charge multiple receivers simultaneously similar to WiFi communications. To guarantee the Quality of Charging Service (QoCS) for each receiver and maximize the overall earning in the PtMP RBC service, we specify the Charging Pricing Strategy (CPS) and develop the High Priority Charge (HPC) scheduling algorithm to control the charging order and power allocation. Each receiver is assigned a priority, which is updated dynamically based on its State of Charging (SOC) and specified charging power. The receivers with high priorities are scheduled to be charged in each time slot. We present the pseudo code of the HPC algorithm based on quantifying the receiver's SOC, discharging energy and various relevant parameters. Relying on simulation analysis, we demonstrate that the HPC algorithm can achieve better QoCS and earning than the Round-Robin Charge (RRC) scheduling algorithm. Based on the performance evaluation, we illustrate that the methods to improve the PtMP RBC service are: 1) limiting the receiver number within a reasonable range and 2) prolonging the charging duration as long as possible. In summary, the HPC scheduling algorithm provides a practical strategy to maximize the earning of the PtMP RBC service with each receiver's QoCS guarantee.
\end{abstract}

\begin{IEEEkeywords}
Wireless charging, quality of charging service, charging pricing strategy, scheduling algorithm
\end{IEEEkeywords}

\IEEEpeerreviewmaketitle

%%%%%%%%%%%%%%%%%%%%%%%%%%%%%%%%%%%%%%%%%%%%%%%%%%%%%%%%%%%%%%%%%%%%%%%%%%%%%%%%%%%%%%%%%%%%%%%%%%%%%%%%%%%%%%%%%%%%
\section{Introduction}\label{Section1}
Benefits from the advance of the Internet of Things (IoT), the interconnection of everything in the world can be realized \cite{gubbi2013internet}. Meanwhile, Qihui Wu and Guoru Ding have developed a new paradigm in \cite{wu2014cognitive}, named Cognitive Internet of Things (CIoT), to empower the current IoT with a ‘brain’ for highlevel intelligence. Thus, an increasing number of objects surrounding us will be connected to the Internet in one form or another, and the intelligent management and unified deployment of objects will be achieved conveniently \cite{ding2018amateur}.

However, the weak battery endurance is a common existing problem for IoT devices. Meanwhile, the high performance computing and communicating lead high requirements for the battery endurance \cite{wang2018power}. Therefore, improving the battery endurance of the devices is vital to IoT development. As an efficient way to prolong the battery run time, the wireless charging (i.e. Wireless Power Transfer, WPT) is developing rapidly recently \cite{hui2014critical, lu2016wireless}.

The traditional wireless charging technologies, including inductive coupling, magnetic resonance coupling, radio frequency, etc., can charge IoT devices conveniently as invested in \cite{wirelesstechniques, electromagnetic}. However, they still face safety, high-power and mobility challenges. To support the high-power, long-distance, safe and mobile charging for IoT devices, Resonant Beam Charging (RBC), as known as Distributed Laser Charging (DLC), is presented in \cite{liu2016dlc}.

\begin{figure}[!t]
	\centering
    \includegraphics[scale=0.2]{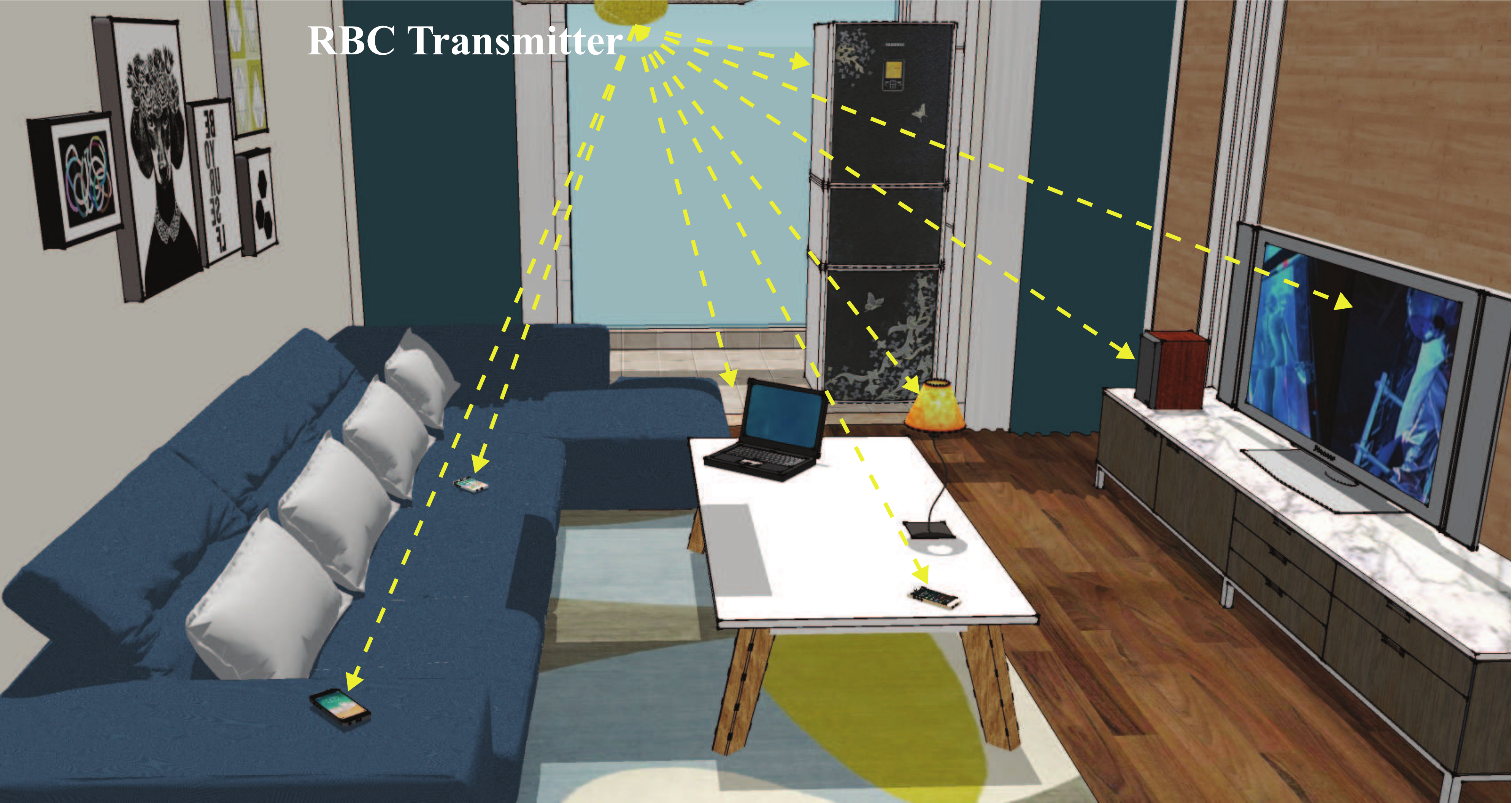}
	\caption{RBC Application Scenario}
    \label{application}
    \end{figure}

Compared with traditional wireless charging technologies, RBC appears to be more suitable for charging mobile IoT devices \cite{Qing2017}. On the one hand, the resonant beam can be generated as long as the transmitter and the receiver are in the Line of Sight (LOS) of each other, and the beam will be cut off when an obstacle enters the LOS. Thus, the mobile and safe wireless charging can be realized in the RBC system. On the other hand, multiple resonant beams can be generated simultaneously, that is, multiple receivers can be charged at the same time. In Fig. \ref{application}, multiple types of receivers embedded with the RBC receiver, e.g. mobile phone, TV, laptop and so on, are charged by a RBC transmitter simultaneously, their State of Charging (SOC, i.e. the battery remaining capacity percentage) and using statuses are different.

Similar to the Policy Control and Charging (PCC) for guaranteeing the value of network resource in wireless communication system, the Charging Pricing Strategy (CPS) should be formulated to ensure the value of charging resource in the RBC system \cite{feder2011mobility, foottit2011system}. What's more, in analogy to the Quality of Service (QoS) for the web service, the Quality of Charging Service (QoCS) should be introduced to quantify the RBC service for each receiver \cite{menasce2002qos, sood2017current}. Since the status, e.g. using status, SOC, specified charging strategy and so on, of each receiver is various, the QoCS for each receiver must be satisfied with different charging power and duration. In addition, the receiver's charging power and duration are also key factors related to the earning. Therefore, guaranteeing each receiver's QoCS and maximizing earning are two crucial issues for the RBC service. To solve these two issues, it is desired to design a reasonable scheduling algorithm to schedule the transmitting power for charging the receivers.

The contributions of this paper include:
1) To guarantee the QoCS for each receiver and maximize the earning in the Point-to-Multipoint RBC service, we propose the High Priority Charge (HPC) scheduling algorithm to schedule the transmitting power.
2) Based on the quantitative analysis of the algorithm implementation, we obtain the closed-form of the relationship between the State of Charging (SOC) and the charging duration under different charging power, and the battery discharging power of each receiver is dynamic depending on different using status.
3) We analyze the performance of the HPC scheduling algorithm, and find that:

\begin{itemize}
  \item Compared with the Round-Robin charge (RRC) scheduling algorithm, the HPC algorithm can guarantee the better QoCS for each receiver and achieve the higher earning.
  \item The methods to improve the performance of the HPC scheduling algorithm include limiting the receiver number within a reasonable range and extending the charging time as long as possible.
\end{itemize}
\begin{figure}[!t]
	\centering
    \includegraphics[scale=0.38]{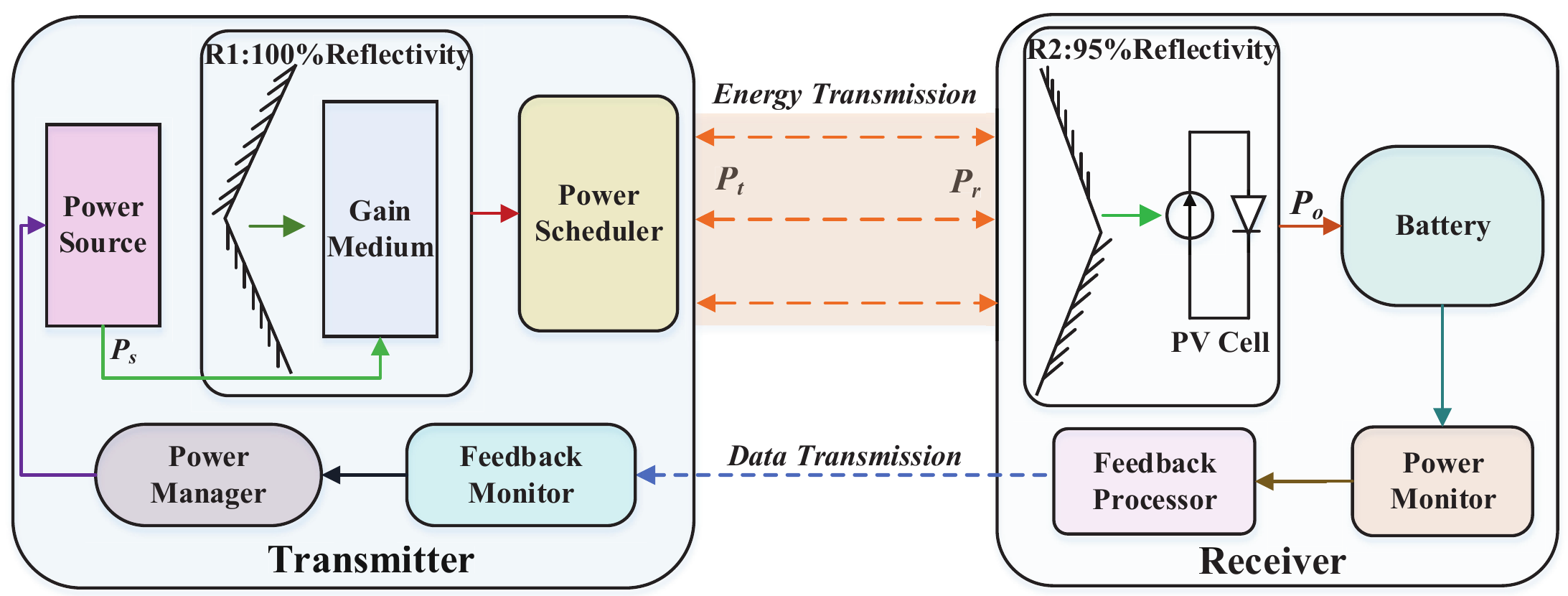}
	\caption{RBC System Structure}
    \label{RBC Structure}
    \end{figure}
In the rest of this paper, we will present the RBC system architecture firstly. In section III, we will design the HPC scheduling algorithm based on the CPS and the receiver status (the SOC and the using status, etc.). In section IV, we will present the algorithm implementation by quantifying the charging-discharging stage and various relevant parameters, and depicting the pseudo code. In section V, we will analyze the performance of the HPC algorithm by simulation. Finally, we will give the summarizing remarks and present the open issues for future research.

%%%%%%%%%%%%%%%%%%%%%%%%%%%%%%%%%%%%%%%%%%%%%%%%%%%%%%%%%%%%%%%%%%%%%%%%%%%%%%%%%%%%%%%%%%%%%%%%%%%%%%%%%%%%%%%%%%%%
\section{RBC System Architecture}\label{Section2}
In this section, we will introduce the structure of the Point-to-Point (PTP) RBC system firstly. Then, the key feature of the RBC system, Point-to-Multipoint (PtMP) charging, will be analyzed.

\subsection{Point-to-Point RBC System}\label{}
Multiple IoT devices can be charged simultaneously in the RBC system while guaranteeing the high-power, long-distance, mobile and safe charging \cite{liu2016dlc}. A Point-to-Point (PTP) RBC system consists of a RBC transmitter and a RBC receiver, of which the structure is presented in Fig. \ref{RBC Structure} \cite{fang2018}.

In Fig. \ref{RBC Structure}, the RBC transmitter includes a power source, a retro-reflector R1 with 100\% reflectivity, a gain medium, a power scheduler, a feedback monitor and a power manager. A retro-reflector R2 with 95\% reflectivity, a PV cell, a battery, a power monitor and a feedback processor consist the receiver. Based on the receiver's charging status (e.g. SOC, charging power, charging time and so on) obtained by the feedback monitor, the power source provides the supplied power $P_s$ for the gain medium to pump resonant beam under the control of the power manager. The resonant beam power $P_t$ scheduled by the scheduler is transmitted to the receiver. The resonant beam power received by the receiver $P_r$ can partially pass through R2, and the beam behind R2 can be converted into the output electrical power $P_o$ by the PV cell to charge battery. The battery charging status is monitored by the power monitor, which will then inform the feedback processor to transfer the status information. The symbols of the PTP RBC system in Fig. \ref{RBC Structure} are listed in Table \ref{mulparameter}.

    \begin{table}[!t]
    \setlength{\abovecaptionskip}{0pt}
    \setlength{\belowcaptionskip}{-3pt}
    \centering
        \caption{RBC System Symbols}
    \begin{tabular}{C{0.6cm} C{7cm}}
    \hline
     \textbf{Symbol} & \textbf{Parameter}  \\
    \hline
    \bfseries{$P_s$} & {Supplied power} \\
    \bfseries{$P_t$} & {Transmitting power (Transmitter beam power)} \\
    \bfseries{$P_r$} & {Receiver beam power } \\
    \bfseries{$P_o$} & {Receiver available charging power (Output electrical power)} \\
    \hline
    \label{mulparameter}
    \end{tabular}
    \end{table}

Since the RBC transmitter and receiver are separated in space, the transmitter can be embedded in the ceiling lamp, the router and so on to provide the transmitting power. The receiving devices, like mobile phone, watch and sensor, can be embedded with the RBC receiver, and charged simultaneously.

\subsection{Point-to-Multipoint RBC System}\label{}

The RBC system which includes a transmitter and multiple receivers can be called Point-to-Multipoint (PtMP) RBC system, where the types of receivers are various, such as mobile phone, laptop, TV, sensor. In addition, the receiver statuses, e.g. the SOC, the specific charging power and the using status, are different. Thus, when multiple charging connections between the transmitter and receivers are established, the charging respond time (i.e. the time from accessing to RBC system to charging), the charging rate (i.e. the charging power) and the charging duration of each receiver are diverse.

On the other hand, the QoCS for each receiver in the PtMP RBC system is related to the receiver's charging respond time and the charging rate. The charging respond time is related to the receiver charging order and the charging rate is decided by the charging power. What's more, the charging power and the charging duration are the critical factors determining the overall earning. Thus, to ensure the efficiency of charging multiple receivers simultaneously in the PtMP RBC system, it is vital to control the receiver charging power, order and duration efficiently. That is, maximizing the earning with guaranteeing the QoCS for each receiver is crucial for the charging process in the PtMP RBC service.

In the PtMP RBC system, to guarantee the receiver's QoCS and maximize the earning, an appropriate scheduling algorithm must be proposed for controlling the receiver charging power, order, duration and so on. In the next section, we will propose the High Priority Charge (HPC) scheduling algorithm to solve the WPT scheduling problem for IoT devices.

%%%%%%%%%%%%%%%%%%%%%%%%%%%%%%%%%%%%%%%%%%%%%%%%%%%%%%%%%%%%%%%%%%%%%%%%%%%%%%%%%%%%%%%%%%%%%%%%%%%%%%%%%%%%%%%%%%%%
\section{HPC Algorithm Design}\label{Section2}

In this section, we will specify the regulations of formulating the Charging Pricing Strategy (CPS), the factors related to the Quality of Charging Service (QoCS) and the earning. On this basis, the High Priority Charge (HPC) scheduling algorithm will be proposed, and its execution flow will be presented.

\subsection{Charging Pricing Strategy}\label{}
Analogous to Policy Control and Charging (PCC), the Charging Pricing Strategy (CPS) of the RBC service can be determined by charging power, charging level, charging duration or other factors. In this paper, the CPS related to the charging power $P_c$ and the charging duration $T_s$ is formulated, and it is depicted in Table \ref{Charge Pricing}.
\begin{table}[!h]
    \setlength{\abovecaptionskip}{0pt}
    \setlength{\belowcaptionskip}{-3pt}
    \centering
        \caption{Charging Pricing Strategy}
    \begin{tabular}{C{1.4cm} C{1.8cm} C{2cm} C{1.8cm}}
    \hline
    \textbf{Priority $P_p$}  & \textbf{Power $P_c$ (W)} & \textbf{Duration $T_s$ (h)}  & \textbf{Pricing $C_p$ (\$)}\\
    \hline
    \bfseries{1} & {5} & {1} & {1}\\
    \bfseries{2} & {10} & {1} & {3}\\
    \bfseries{3} & {15} & {1} & {5}\\
    \hline
    \label{Charge Pricing}
    \end{tabular}
    \end{table}

For the CPS in Table \ref{Charge Pricing}, the charging power $P_c$ is 5W, 10W and 15W respectively, and the corresponding pricing $C_p$ with 1 hour is \$1, \$3 and \$5 respectively. In addition, we assume that the priority $P_p$ of the three charging strategy 5W, 10W, 15W is 1, 2 and 3.

\subsection{Quality of Charging Service}\label{}
The Quality of Charging Service (QoCS) for each RBC receiver is determined by the charging response time, the charging rate, the SOC and the charging duration, etc. In this paper, we assume that the QoCS depends on the following factors:

1) The charging response time, that is the time from the receiver accessing to the RBC system to starting being charged, which is decided by the receivers charging order.

2) The charging rate, i.e. the receiver charging power corresponding to the specified charging strategy.

3) The charging duration of each receiver can be sustained during a charging process.

Therefore, all factors influence the QoCS are the charging parameters during a charging process, and the QoCS for each receiver can be reflected by the State of Charging (SOC).

\subsection{Earning}\label{}

During the charging process in the PtMP RBC system, the earning is related to the CPS and the receiver QoCS. Thus, the earning $E_r$ in a charging process depends on the charging duration of different specified charging strategies $T_e$ and the charging pricing $C_p$ (see Table \ref{Charge Pricing}). The relationship among them is depicted as:
\begin{equation}\label{charge-earning}
    E_{r} = \sum_{P_p = 1}^{3} T_e \cdot C_p,
\end{equation}
where the charging power $P_c$ corresponding to the priority $P_p$ 1, 2, and 3 is 5W, 10W and 15W respectively, and the value of $C_p$ is \$1, \$3 and \$5 correspondingly. $E_r$ can be reflected by the charging financial gains directly.

\subsection{HPC Algorithm}\label{}
In the PtMP RBC system, guaranteeing the receiver's QoCS and maximizing the earning can be realized by scheduling the transmitting power reasonably. The High Priority Charge (HPC) scheduling algorithm is proposed to complete the reasonable scheduling.

\begin{itemize}
  \item Design Ideas
\end{itemize}

The design ideas of the HPC scheduling algorithm include: 1) The charging time is divided into multiple equal charging time slots. 2) All receivers are queued to a receiver queue from high to low priority. 3) A few receivers in head of the queue can be charged simultaneously with the specified charging power (5W, 10W or 15W). 4) All receivers discharge depending on their using statuses during the charging process.

The receiver priority is determined by the SOC and the specified charging strategy, which is higher with the lower SOC $C_t$ (unit is \%) and higher specified charging priority $P_p$ (i.e., higher charging power $P_c$). Thus, the receiver priority $R_p$ of the receiver $i$ can be described as:
\begin{equation}\label{charge-priority}
    R_{p}(i) = (1-C_t(i)) \times P_p(i).
\end{equation}

\begin{itemize}
  \item Execution Flow
\end{itemize}

Based on the design ideas, the execution flow of the HPC algorithm is depicted in Fig. \ref{HPCflowchart}. The detailed descriptions of each execution step are presented as follows:

\begin{figure}[!t]
	\centering
    \includegraphics[scale=0.6]{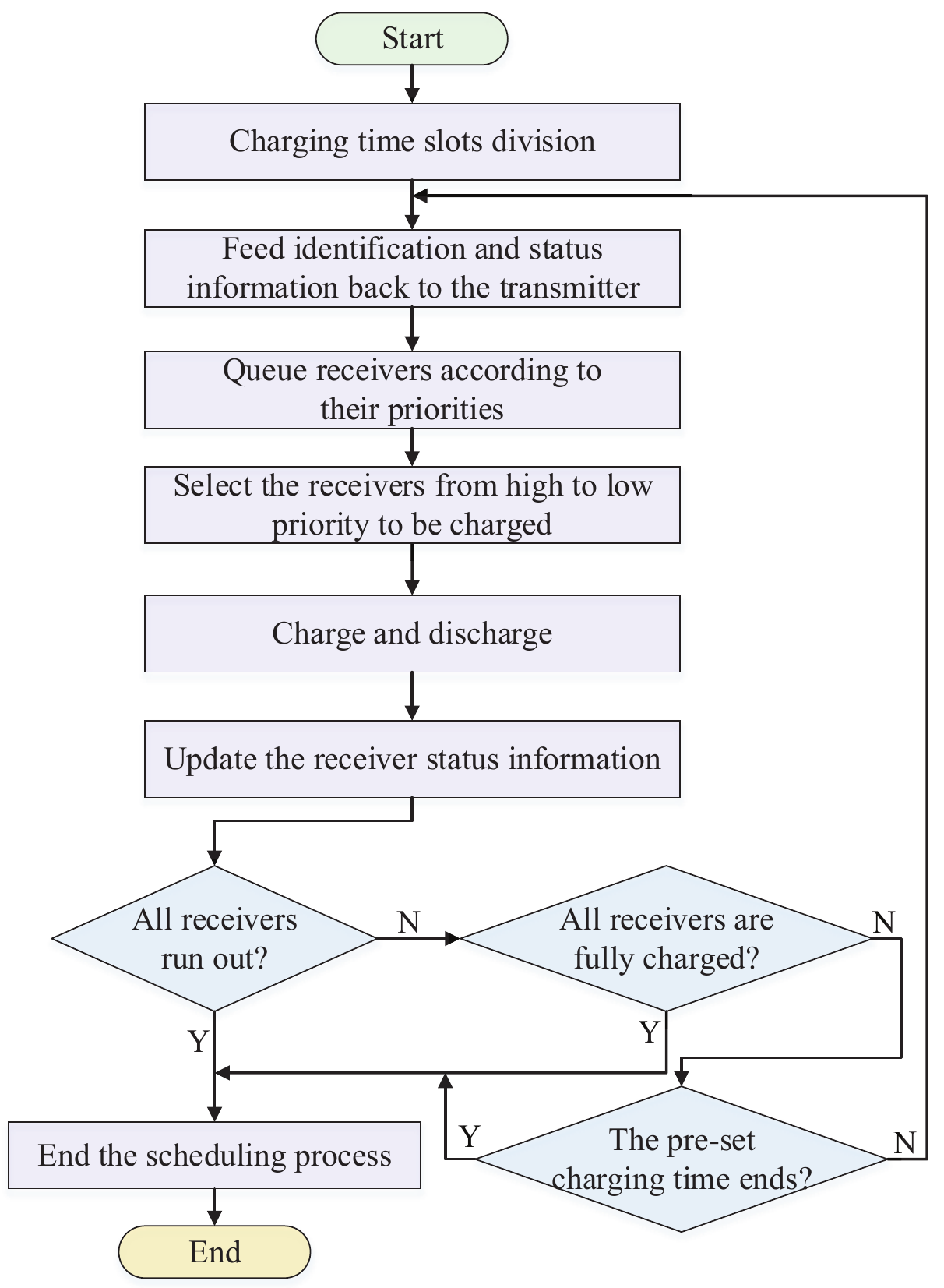}
	\caption{HPC Scheduling Algorithm Flow Chart}
    \label{HPCflowchart}
    \end{figure}

1) Time division: The charging time is divided into multiple equal time slots, and the duration of each time slot can be determined according to the specific charging status.

2) Feedback: The identifications and status information of the receivers are fed back to the transmitter. The status information mainly includes the receiver specified charging strategy and the SOC, etc.

3) Queueing: All receivers are queued to a receiver queue according to their priorities from high to low. The priority can be determined by the specified charging strategy and the SOC based on \eqref{charge-priority}.

4) Selecting: The receivers with higher priority in head of the queue are selected to be charged firstly.

5) Charging and discharging: The selected receivers are charged with the power corresponding to the specified charging strategy, and all receivers discharge depending on their using statuses.

6) Updating: The receiver's SOC is updated after charging and discharging.

7) End judgement: The charging process ends when one of the following three condition is satisfied: a) all receivers run out, b) all receivers are fully charged, and c) the pre-set charging time ends. Otherwise, turn to 2).

\begin{figure}[!t]
	\centering
    \includegraphics[scale=0.48]{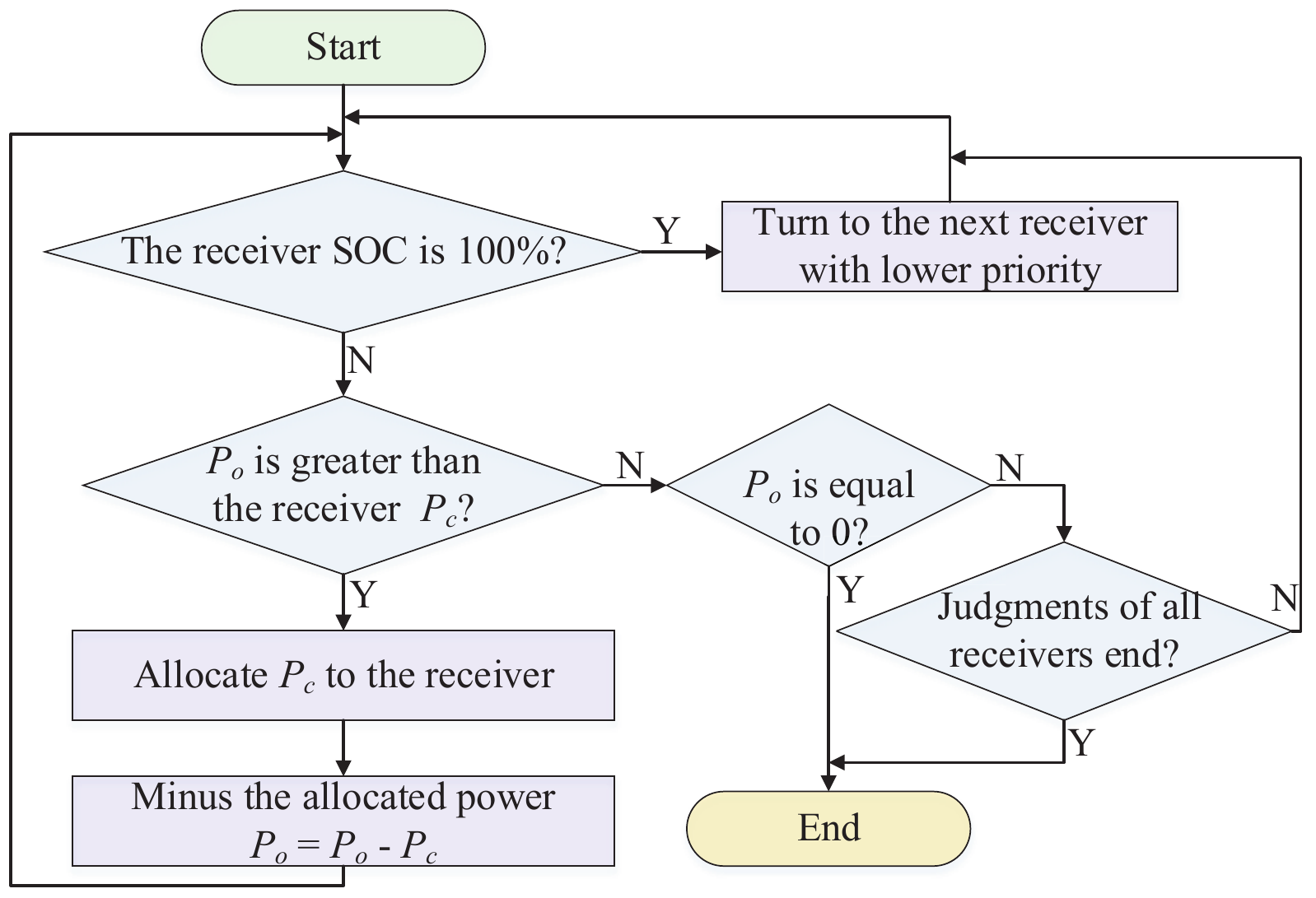}
	\caption{HPC Transmitting Power Allocation}
    \label{allocatestep}
    \end{figure}

In addition, allocating the transmitting power $P_t$ to charge the receivers is a key step in a charging process. The allocation process is shown in Fig. \ref{allocatestep}, and $P_o$ is the receiver available charging power decided by $P_t$, $P_c$ is the charging power corresponding to the receiver specified charging strategy.

From Fig. \ref{allocatestep}, the transmitting power is allocated to the receivers one by one in accordance with the priority from high to low. When the available charging power $P_o$ is greater than the charging power $P_c$ specified by the selected receiver, the receiver can be charged with $P_c$. Otherwise, the judgment is continued until the transmitting power is fully allocated or all receivers are traversed.

In this section, the HPC scheduling algorithm is proposed. To implement the HPC algorithm, it's necessary to analyze each charging step quantitatively. Moreover, various parameters related to the algorithm implementation (e.g., the receiver number, the charging efficiency and so on) need to be quantitatively specified. In the next section, we will quantitatively analyze the variables and processes involved in the algorithm implementation.
%%%%%%%%%%%%%%%%%%%%%%%%%%%%%%%%%%%%%%%%%%%%%%%%%%%%%%%%%%%%%%%%%%%%%%%%%%%%%%%%%%%%%%%%%%%%%%%%%%%%%%%%%%%%%%%%%%%%
\section{HPC Algorithm Implementation}\label{Section2}

In this section, we will propose the quantitative analysis methods of the charging and discharging stages, specify the parameters involved in the implementation quantitatively, and depict the implementation pseudo code.

\subsection{Charging Quantification}\label{}

According to the established CPS in the above section, the charging power specified by the receivers is 5W, 10W and 15W respectively, of which the occurrence frequency is uniformly distributed when multiple receivers are charged simultaneously.

The charging power $P_{c}$ of the CPS is:
\begin{equation}\label{charge-power}
    P_{c}=\{5W\  10W\  15W\}.
    \end{equation}

The occurrence frequency $F_{p}$ of the charging power 5W, 10W and 15W is:
\begin{equation}\label{charge-rate}
    F_{p}=\{\frac{1}{3}\  \frac{1}{3}\  \frac{1}{3}\}.
    \end{equation}

The charging power $P_{c}$ of a receiver during a charging process is:
  \begin{equation}\label{charge-power-each}
        \begin{aligned}
        P_{c}=randsrc(1,1,[P_{r};F_{p}]).
        \end{aligned}
        \end{equation}

\begin{figure}[!t]
	\centering
    \includegraphics[scale=0.55]{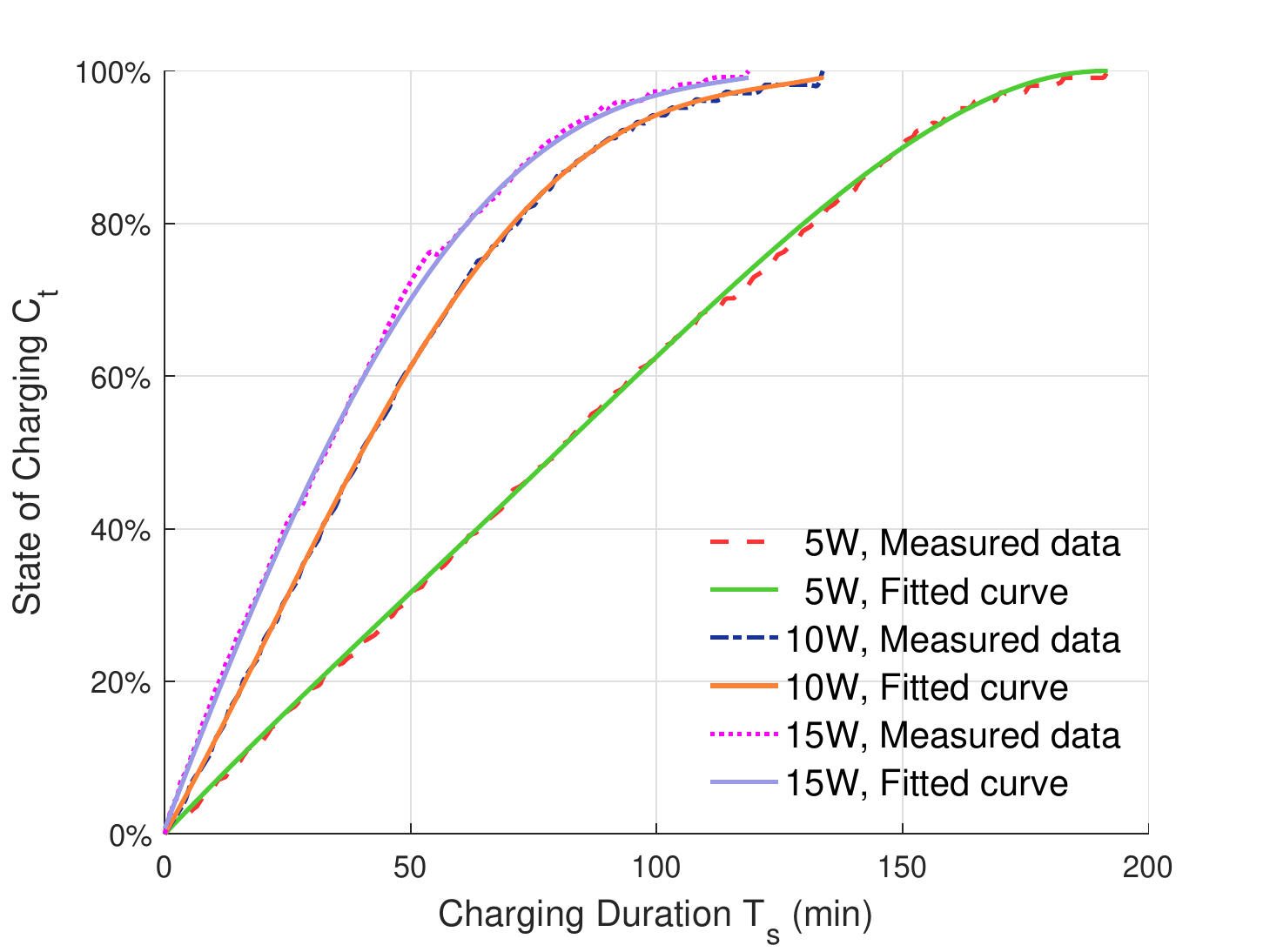}
	\caption{State of Charging vs. Charging Duration}
    \label{chargetime-chargecapacity}
    \end{figure}

In addition, the charging power and the charging duration are determinants of the SOC. To obtain the specific relationship between them, we investigate the measured data and fit the relationship of them. As a kind of widely used IoT device, we take iPhone8 plus as an example for analysis, whose battery capacity and energy are 2691mAh, 10.28Wh respectively \cite{Iphone8-review, Iphone8-review2}. According to the measured data of State of Charging (SOC) $C_t$ with the charging duration $T_s$ under different charging power $P_c$ in \cite{Iphone8, Iphone82}, we plot the SOC and the charging duration in Fig. \ref{chargetime-chargecapacity}.

In Fig. \ref{chargetime-chargecapacity}, the measured data of SOC with 5W is shown as the dash line, 10W is the dotted-dash line and 15W is the dotted line. The SOC increases over the charging duration, and the growth rate of 15W is greater than that of 10W and 5W. Therefore, the time of charging fully with 15W is shorter than 10W and 5W, and the time is 118min, 130min and 190min approximately.

To obtain the SOC $C_t$ from the charging duration $T_c$ directly, we fit the relationship of them in the MATLAB curve-fitting toolbox. The fitted function is a quartic polynomial as:
\begin{equation}\label{chargecapacity-time}
{C}_{t}= a T_c^4 + b T_c^3 + c T_c^2 + d T_c + e.
\end{equation}

The values of the fitting coefficients are shown in Table \ref{Fitting Coefficients}. The fitted curves of 5W, 10W, 15W are depicted as the lines in Fig. \ref{chargetime-chargecapacity}, and the fitted curve can almost completely fit the measured data.
\begin{table}[!t]
    \setlength{\abovecaptionskip}{0pt}
    \setlength{\belowcaptionskip}{-3pt}
    \centering
        \caption{Fitting Coefficients}
    \begin{tabular}{C{1.5cm} C{1.8cm} C{1.8cm} C{1.8cm}}
    \hline
     \textbf{Coefficient} & \textbf{5W}  & \textbf{10W} & \textbf{15W} \\
    \hline
    \bfseries{$a$} & { -7.858e-10 } & { 5.372e-09 } & {3.885e-09} \\
    \bfseries{$b$} & { 1.992e-07 } & {-1.561e-06} & {-8.323e-07} \\
    \bfseries{$c$} & { -1.764e-05 } & { 8.361e-05 } & {-2.837e-05} \\
    \bfseries{$d$} & { 0.006813 } & {0.0113} & {0.01689} \\
    \bfseries{$e$} & { 0 } & {0} & {0.006894}\\
    \hline
    \label{Fitting Coefficients}
    \end{tabular}
    \end{table}

\subsection{Discharging Quantification}\label{}
The discharging power of each receiver is decided by its type and using status. We take the mobile phone as the study object in this paper, whose discharging power depends on the using status \cite{fang2018, carroll2010analysis}. Thus, we study the using statuses and discharging power of mobile phones in this subsection.

\begin{figure}[!t]
	\centering
    \includegraphics[scale=0.4]{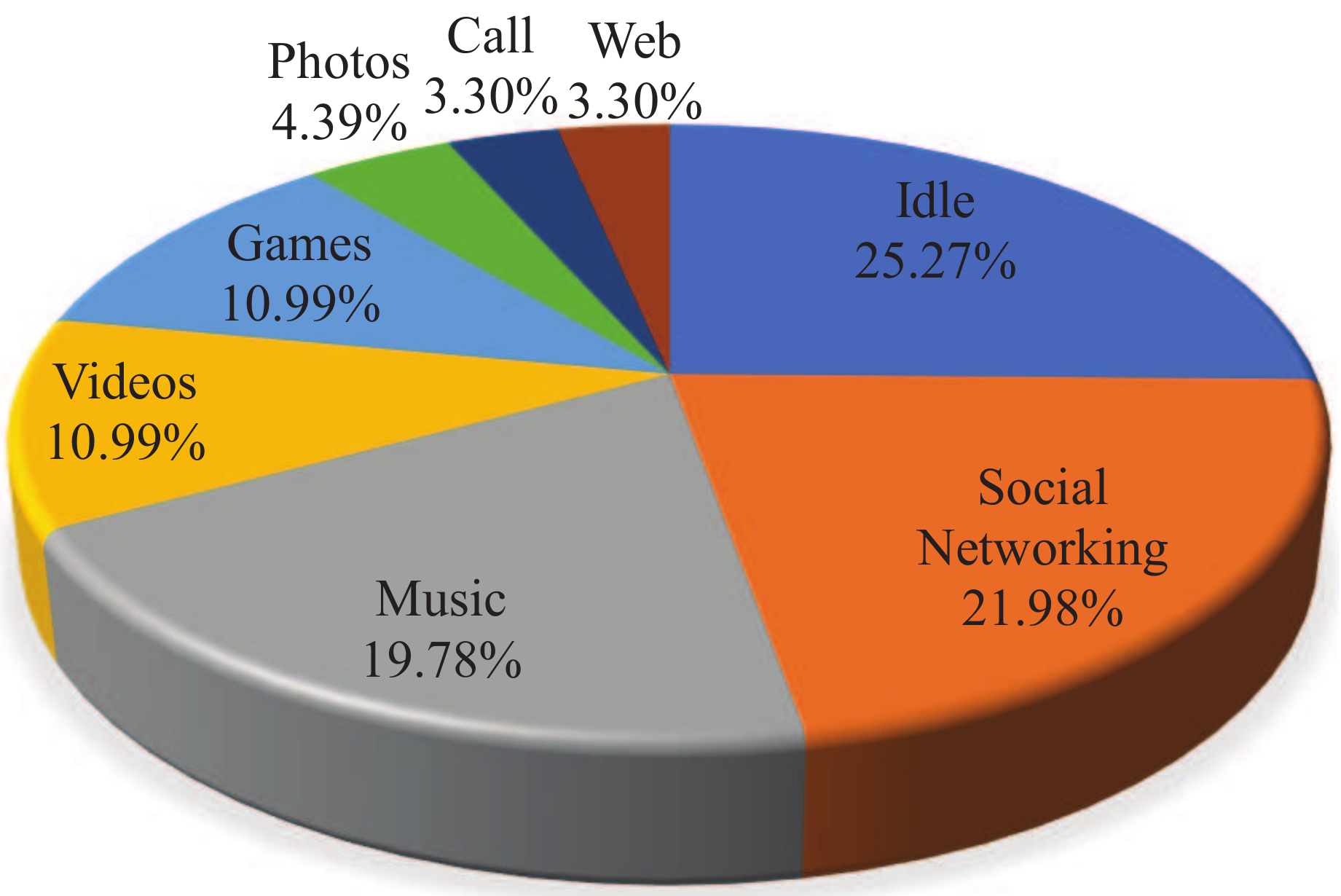}
	\caption{Using Status vs. Using Rate}
    \label{usingstates-rate}
    \end{figure}

\begin{figure}[!t]
	\centering
    \includegraphics[scale=0.43]{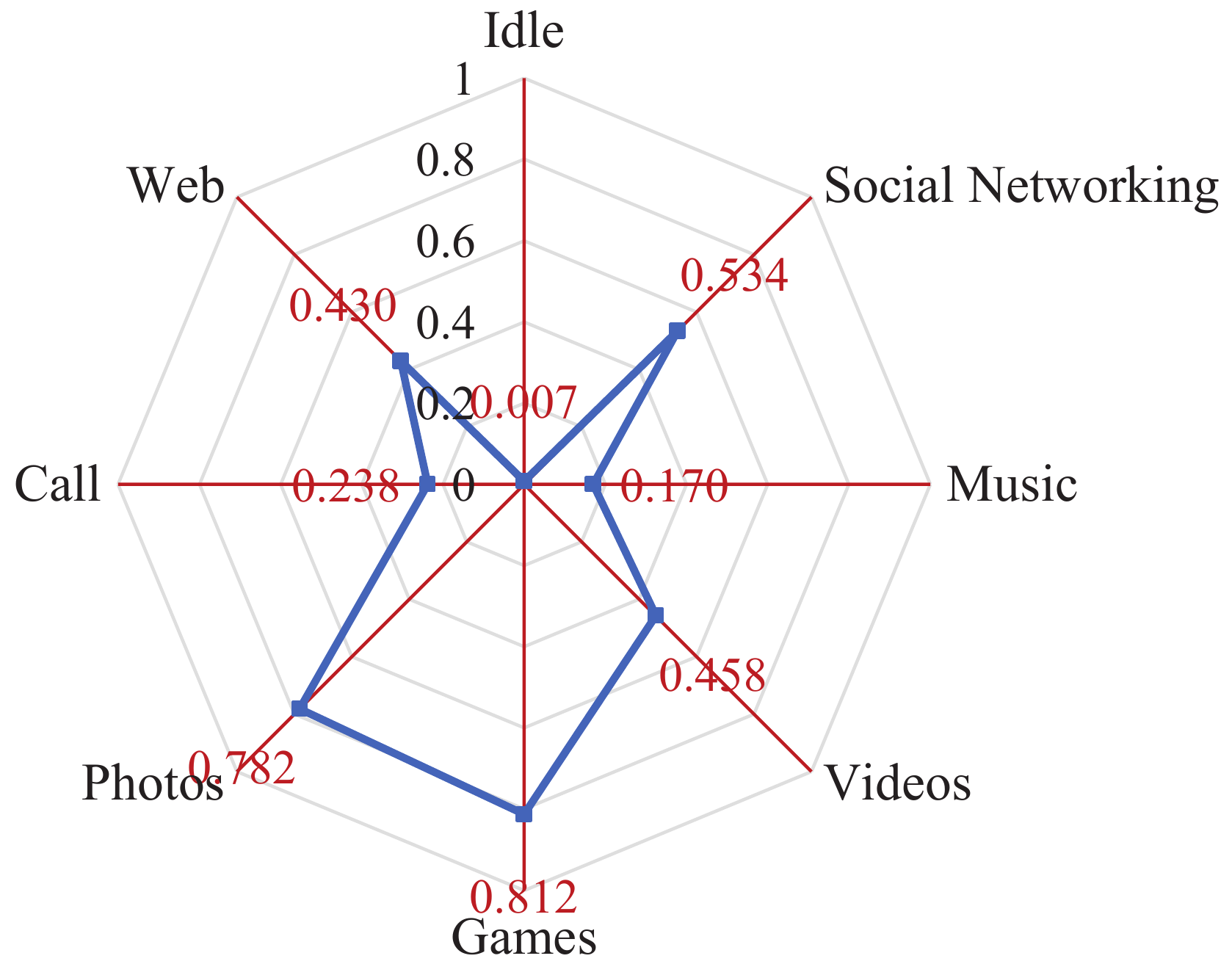}
	\caption{Discharging Power vs. Using Status}
    \label{usingstate-dischargepower}
    \end{figure}

Based on the investigations, the frequently-used statuses and their using rates of the mobile phone are presented in Fig. \ref{usingstates-rate} \cite{carroll2010analysis, murmuria2012mobile}. From Fig. \ref{usingstates-rate}, the using statuses of the mobile phone are idle, social networking, music, videos, games, phones, call, and web. The most commonly using status is idle, followed by social networking. In addition, the discharging power of each using status is depicted in Fig. \ref{usingstate-dischargepower}.

From Fig. \ref{usingstate-dischargepower}, the discharging power of different using status is different. The highest discharging power is games, and photos is the second. The discharging power is lowest when the mobile phone is idle. Thus, the discharging power of each receiver at each charging time slot can be quantized as:

The using rates $U_{r}$ of idle, social networking, music, videos, games, phones, call and web are: \begin{equation}\label{usage-rate}
\begin{aligned}
    U_{r}=\{25.27\%\  21.98\%\  19.78\%\  10.99\%\\\
       10.99\%\   4.39\%\   3.30\%\   3.30\%\}.
\end{aligned}
\end{equation}

The unit of discharging power $P_{u}$ is W, and the values of common using statuses are:
\begin{equation}\label{discharge-power}
    P_{u}=\{0.007\  0.534\  0.170\  0.458\   0.812\   0.782\   0.238\   0.430\}.
    \end{equation}

The discharging power $P_{d}$ of each receiver during a charging time slot is:
  \begin{equation}\label{discharge-power-each}
        \begin{aligned}
        P_{d}=randsrc(1,1,[P_{u};U_{r}]).
        \end{aligned}
        \end{equation}
Moreover, the discharging energy is the integral of the discharging power over the charging time:
\begin{equation}\label{discharge-energy}
    E_d = \int_{0}^{T_s} P_d(t)\,dt,
    \end{equation}
where $T_s$ is the charging duration, i.e. the discharging duration \cite{fang2018}. The percentage of the power consumption $D_p$ over a period of time is the ratio of the consumed energy over the time $E_d$ to the battery total energy $E_t$, and it is depicted as:
\begin{equation}\label{discharge-percentage}
    D_p = \frac{E_d}{E_t},
    \end{equation}
where the total energy of the iPhone8 plus battery is 10.28Wh, that is $E_t=10.28$.

\subsection{Implementation Pseudo Code}\label{}
In the above subsections, we have quantified the most important stages in the charging process: the charging stage and the discharging stage. However, to fully implement the HPC scheduling algorithm, some relevant parameters need to be specified quantitatively. The settings of each parameter are as follows:

1) Receiver Initial State of Charging

Before the charging process starts, the initial SOC of each receiver is a random number between 0\% and 100\%. The initial SOC $C_{tini}$ of the receiver $R_i$ can be depicted as:
\begin{equation}\label{remaining-capacity}
    \begin{aligned}
    C_{tini}(R_i)=randi([0,100],1,1)\%.
    \end{aligned}
    \end{equation}

2) Receiver Number

In the PtMP RBC system, multiple receivers can be charged simultaneously. During a scheduling process, the receiver number $N_r$ is a random number.

3) Charging Efficiency

In the PTP RBC system depicted in Fig. \ref{RBC Structure}, a charging process includes the electro-optical conversion, the transmitting power scheduling, the beam transmission, the photoelectric conversion, the charging power monitor, the information feedback, and the power management. Among them, the electro-optical conversion efficiency ${\eta}_{el}$, the beam transmission efficiency ${\eta}_{lt}$, and the photoelectric conversion efficiency ${\eta}_{le}$ are the primary factors that affect the overall charging efficiency ${\eta}_{o}$ \cite{Qing2017}. ${\eta}_{o}$ is the result of multiplying ${\eta}_{el}$, ${\eta}_{lt}$ and ${\eta}_{le}$:
\begin{equation}\label{etao}
    {\eta}_{o}={\eta}_{el} {\eta}_{lt} {\eta}_{le}.
    \end{equation}

In addition, ${\eta}_{o}$ varies with the changes of the influencing factors, e.g. the resonant beam wavelength, the transmission environment and so on. In this paper, we assume that the HPC scheduling algorithm is applied to a specific PtMP RBC system, so the efficiency of each charging stage is fixed and can be assumed as \cite{zhang2018distributed2}:

\begin{itemize}
      \item The electro-optical conversion efficiency ${\eta}_{el}$ is 40\%.
      \item The transmission efficiency ${\eta}_{lt}$ is 100\%.
      \item The photoelectric conversion efficiency ${\eta}_{le}$ is 50\%.
    \end{itemize}

    Thus, the charging efficiency ${\eta}_{o}$ is 20\%  calculated by 40\%$\times$100\%$\times$50\%.

4) Transmitting Power

The transmitting power $P_t$, which is determined by the supplied power $P_s$ and the electro-optical conversion efficiency ${\eta}_{el}$, is the optical power of the transmitter. Since the supplied power $P_s$ and the efficiency ${\eta}_{el}$, ${\eta}_{lt}$, ${\eta}_{le}$ are fixed, the transmitting power $P_t$ and the receiver available charging power $P_o$ are also fixed. The relationship among $P_s$, $P_t$, $P_r$, and $P_o$ depicted in Fig. \ref{RBC Structure} is formulated as:
\begin{equation}\label{transpower}
    \begin{aligned}
    P_{o} = P_s {\eta}_{o} = P_s {\eta}_{el} {\eta}_{lt} {\eta}_{le} =  P_t {\eta}_{lt} {\eta}_{le} = P_r {\eta}_{le}.
    \end{aligned}
    \end{equation}

5) Charging Time Slot

The charging time is divided into multiple equal charging time slots, and the scheduling process is implemented per time slot. A time slot $T_c$ can be determined by the charging characters of the battery (e.g., the charging power, the charging capacity) and the actual charging conditions (such as the receiver number, the charging duration).
\begin{algorithm}[!h]
    \caption{HPC Scheduling Algorithm}
    \begin{algorithmic}[1]
    \Require $N_r$, $C_{rini}$;
    \State initialize $T_{c}$, $T_{p}$, $T_{s} \leftarrow 0$, $T_e \leftarrow 0$;
    \State $R(:,1) \leftarrow C_{rini}$;
    \State $R(:,2) \leftarrow randsrc(1,N_r,[P_r;F_p])$;
    \While {$ R(:,1) \geqslant 0\%$ \textbf{and} $R(:,1) < 100\%$ \textbf{and} $T_s \leqslant T_p$}
    \State initialize $P_t$;
    \State $P_o \leftarrow P_t \times 100\% \times 50\%$;
    \State $R(:,3) \leftarrow randsrc(1,N_r,[P_u;U_p])$;
    \State $R(:,4) \leftarrow (1-R(:,1)) \times P_p$;
    \State $R_{sort} \leftarrow sortrows(R, -4);$  %-4按第四列降序排列
    \State $R_{sort}(:,5) \leftarrow 0$;
    \For {$i \leftarrow 1\ to\ N_r$}
    \If {$R_{sort}(i,1)=100\%$} $R_{sort}(i,5) \leftarrow 1$;
    \EndIf
    \EndFor
    \For {$i \leftarrow 1\ to\ N_r$}
    \If {$R_{sort}(i,5)=0\  \textbf{and}\  P_o \geqslant R_{sort}(i,2)\  \textbf{and}\  P_o > 0$}
    \State $f(x)\leftarrow a x^4 + b x^3 + c x^2 + d x + e - R_{sort}(i,1)$;
    \State $t \leftarrow fzero(f(x),100)$;
    \State $R_{sort}(i,1) \leftarrow a (t+T_c)^4 + b (t+T_c)^3 + c (t+T_c)^2 + d (t+T_c) + e$;
    \State $P_o \leftarrow P_o- 5(|10|15)$;
    \State $T_e(1|2|3) \leftarrow T_e(1|2|3) + T_c$
    \EndIf
    \EndFor
    \State $D_p \leftarrow \frac{R_{sort}(:,3) \times T_c}{E_t} $
    \State $R_{sort}(:,1) \leftarrow R_{sort}(:,1) - D_p;$
    \State $R \leftarrow R_{sort}$
    \State $T_s \leftarrow T_s + T_c$;
    \EndWhile
    \State $E_{r} \leftarrow \sum_{P_p = 1}^{3} T_e \cdot C_p$
    \State \Return{$R(:,1)$, $E_r$};
    \label{HPC}
    \end{algorithmic}
    \end{algorithm}

Therefore, all parameters related to the scheduling process have been defined and specified. Based on the execution flow and the relevant quantitative parameters of the HPC scheduling algorithm, we can depict the pseudo code to implement the algorithm. The pseudo code is written in Algorithm 1, where $R$ is the receiver array, while $R_{sort}$ is the receiver array queued from high priority to low. Different columns in $R$ and $R_{sort}$ have different meanings, e.g., the first column is the receiver SOC and the second column is the specified charge power.

As presented in the pseudo code, the specific implementation processes of the HPC scheduling algorithm are as follows:

1) The receiver number $N_r$ and the receiver initial SOC $R(:,1)$ %($C_r$)
are known.

2) Set the charging time slot $T_c$, the execution time $T_p$, the total charging duration $T_s$ and the charging duration of different charging strategy $T_e$.

3) The charging power $R(:,2)$ %($P_c$)
specified by each receiver is determined based on \eqref{charge-power-each}.

4) When the SOC of all receivers is between 0\% and 100\%, or the pre-set charging time $T_p$ is not reached, a HPC scheduling process is performed. Otherwise, turns to 11).

5) Set the transmitting power $P_t$, and the receiver available charging power $P_o$ is equal to $P_t \times 100\% \times 50\%$.

6) Determine the priority $R(:,4)$ %($R_p$)
of the receivers according to the SOC and the specified charging strategy based on \eqref{charge-priority}, and queue the receivers according to their priorities from high to low.

7) When the available charging power is greater than the charging power specified by a receiver with high priority and the receiver is not fully charged ($R(:,5) = 0$), the receiver will be charged according to the specified charging power.

8) The available charging power minus the corresponding allocated value, and the charging duration of different charging strategy adds a charging time slot.

9) Repeat 7), 8) until all receivers have been traversed or the total available charging power is fully allocated.

10) Calculate the power consumption during a charging time slot based on the discharging power $R(:,3)$ and \eqref{discharge-percentage}, and update the receiver SOC and the charging duration, then turns to 4).

11) Calculate the earning based on \eqref{charge-earning}.

12) Return the receiver SOC $R(:,1)$ and the earning $E_r$ when the scheduling process completes.

In this section, the charging and discharging stages are quantified, the relevant parameters are specified, and the implementation pseudo code as well as steps are presented. Based on the circulation of the scheduling, the receiver SOC and the earning can be obtained after a certain period of charging duration under the fixed transmitting power.

%%%%%%%%%%%%%%%%%%%%%%%%%%%%%%%%%%%%%%%%%%%%%%%%%%%%%%%%%%%%%%%%%%%%%%%%%%%%%%%%%%%%%%%%%%%%%%%%%%%%%%%%%%%%%%%%%%%%
\section{Performance Analysis}
In this section, we will analyze the performance of HPC scheduling algorithm based on the simulation results. The performance analysis will be done in two aspects: the receiver's QoCS analysis and the earning analysis. Based on the principles proposed in section III, the QoCS can be reflected in the receiver SOC (remaining capacity percentage) after charging a period of duration, and the financial gains is a direct indication of the earning.
 \begin{figure}[!t]
	\centering
    \includegraphics[scale=0.6]{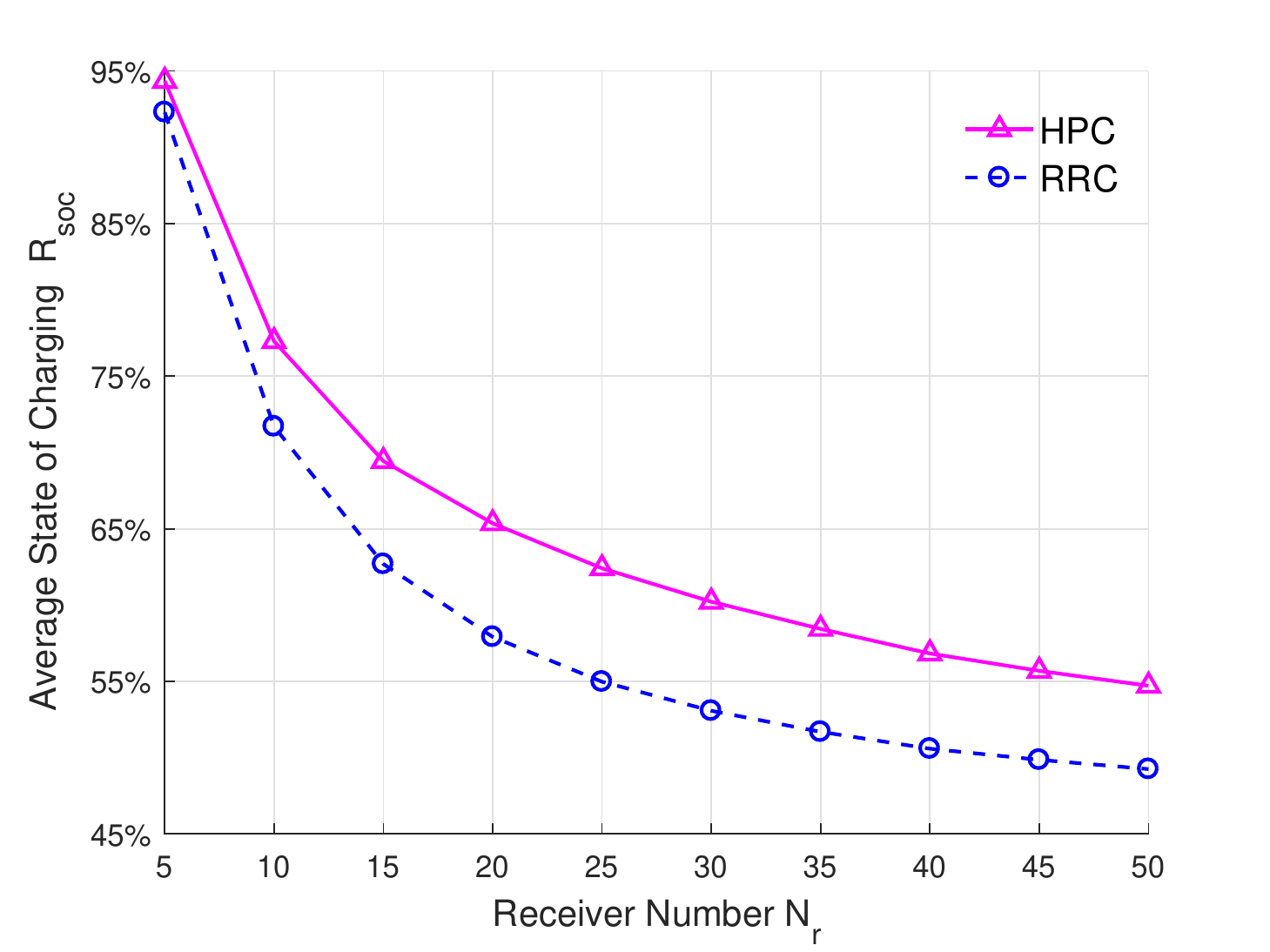}
	\caption{Average State of Charging vs. Receiver Number}
    \label{HPC-roundrobin-cpa}
    \end{figure}
To analyze the performance of HPC scheduling algorithm effectively, we carry out the simulation from two angles: 1) We compare the receiver's QoCS and the earning of the HPC algorithm and the Round-Robin Charge (RRC) scheduling algorithm to analyze the superiority of the HPC algorithm. 2) We analyze the receiver's QoCS and the earning of the HPC scheduling algorithm with different influence factors.

In the simulations, the transmitting power of the PtMP RBC system is assumed as 100W, thus the total available charging power for the receivers is 50W calculated by $100\times100\%\times50\%$. Besides, we assume that the charging time slot is 10s. Since some parameters (e.g. the receiver initial SOC, the discharging power, the specified charging strategy and so on) of each receiver are random during a scheduling process, the ``averaging multiple experiments" method is adopted to eliminate randomness.

\subsection{Comparison}\label{}
To highlight the advantages of HPC scheduling algorithm in the PtMP RBC system, we compare it with the Round-Robin Charge (RRC) scheduling algorithm. The RRC algorithm is easy to implement in the PtMP RBC system and its scheduling principles are as follows:

1) The charging time is divided into multiple equal time slots similar to HPC algorithm.

2) The receivers are queued according to the order of accessing to the system.

3) During a charging time slot, multiple receivers in head of the receiver queue can be charged with their specified charging power until the transmitting power are fully allocated or the judgements of all receivers ends. In addition, all receivers discharge depending on their different using statuses.

4) At the end of a charging time slot, all receivers update their SOC, and the receivers have been charged are queued to tail of the receiver queue.

5) Loop execution 3) and 4) until all receivers are charged fully, or all receivers run out, or the pre-set charging time is reached.

The comparisons of the receiver's QoCS and the earning after charging 2 hours are depicted as follows.

\begin{itemize}
  \item QoCS Comparison
\end{itemize}

    \begin{figure}[!t]
	\centering
    \includegraphics[scale=0.6]{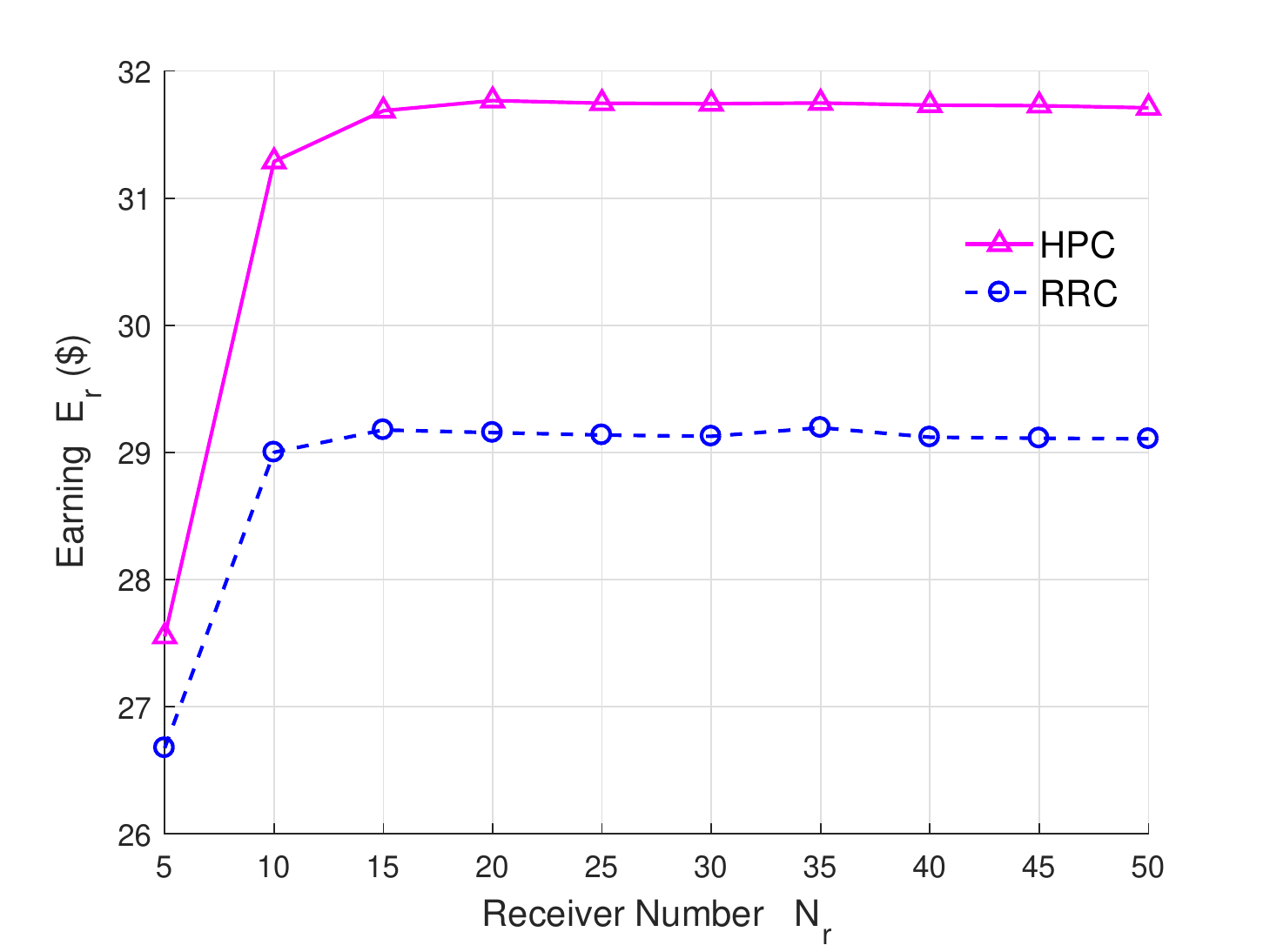}
	\caption{Earning vs. Receiver Number}
    \label{HPC-roundrobin-earn}
    \end{figure}

The comparison of the receiver average SOC $R_{soc}$ with different receiver number is depicted in Fig. \ref{HPC-roundrobin-cpa}. Since the transmitting energy is fixed while the total discharging energy augments as the receiver number increases, the average SOC of the two algorithms decreases with the increasement of receiver number in Fig. \ref{HPC-roundrobin-cpa}. In addition, when the receiver number is same, the average SOC of the HPC algorithm is greater than that of the RRC algorithm. Thus, the QoCS of the HPC scheduling algorithm is better than that of the RRC scheduling algorithm.

\begin{itemize}
  \item Earning Comparison
\end{itemize}

The earning of PtMP RBC service is determined by the CPS and the charging duration. The earning comparison of the two scheduling algorithms is presented in Fig. \ref{HPC-roundrobin-earn}. The transmitting power can not be fully allocated when the receiver number is small, while the allocated transmitting power augments with the receiver number increasing. From Fig. \ref{HPC-roundrobin-earn}, the earning increases rapidly at first, then increases gradually and turns to be constant finally. When increasing the receiver number to a certain value, the transmitting power is fully distributed and the earning reaches the maximum. Due to the randomness of the power distribution, the earning in the curve smooth zone is fluctuate slightly. In addition, the earning of the HPC algorithm is always greater than that of the RRC algorithm in Fig. \ref{HPC-roundrobin-earn}.

As a conclusion, based on the QoCS comparison and the earning comparison, the charging performance of the HPC scheduling algorithm is better than that of the RRC algorithm. Thus, the HPC scheduling algorithm is a superior method to deal with multi-receiver charging in the PtMP RBC system.
     \begin{figure}[!t]
	\centering
    \includegraphics[scale=0.6]{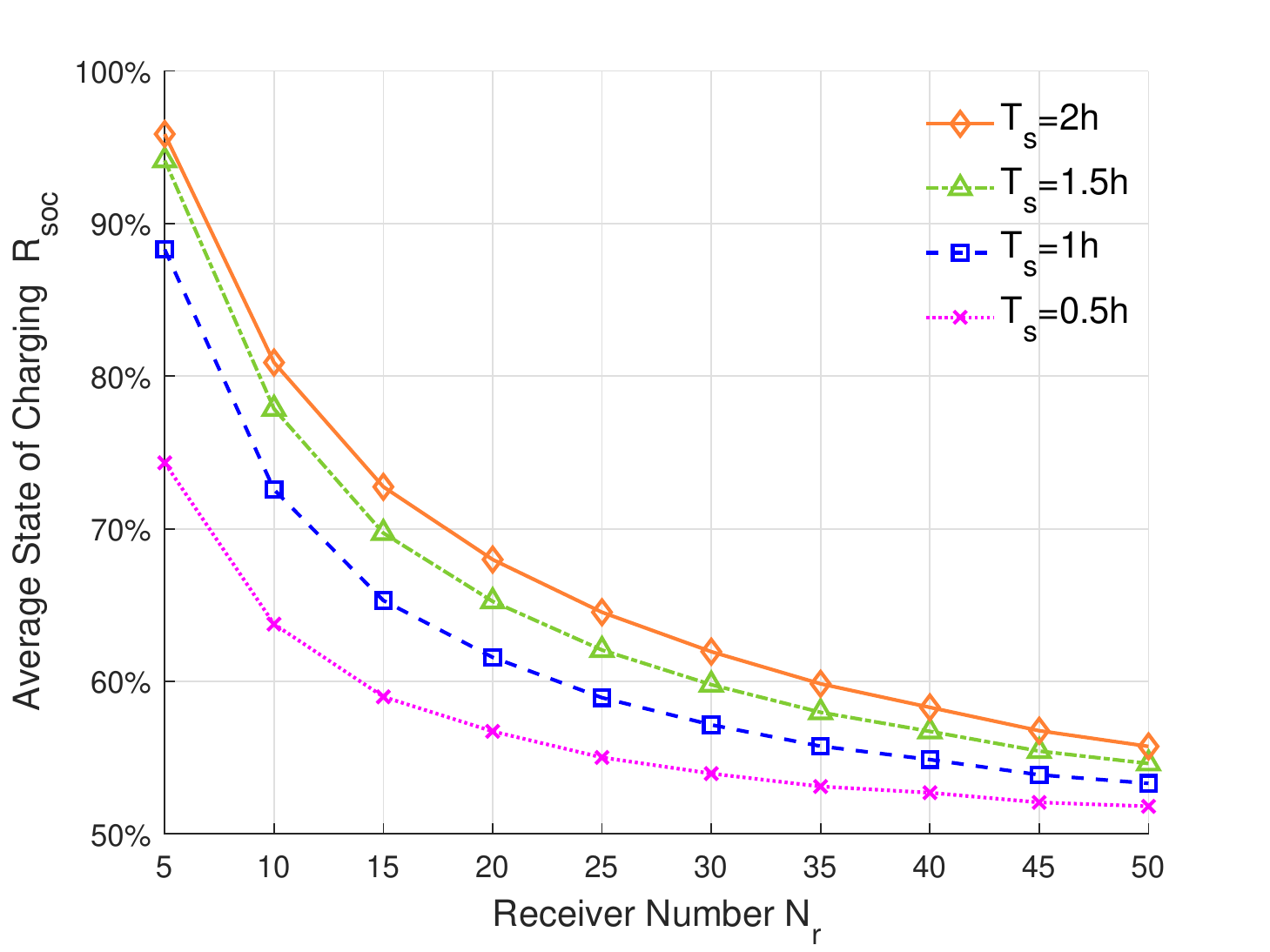}
	\caption{Average State of Charging vs. Receiver Number (Different Charging Duration)}
    \label{chargetime-capacity-time}
    \end{figure}
\subsection{Evaluation}\label{}
In the PtMP RBC system, the performance of the HPC scheduling algorithm varies with different influence factors. In this subsection, we will analyze the performance with different charging duration and receiver number.

\begin{itemize}
  \item QoCS Analysis
\end{itemize}

The QoCS can be reflected by the receiver SOC in a charging process, and the changes of the receiver average SOC $R_{soc}$ with different receiver numbers after charging 0.5h, 1h, 1.5h and 2h are shown in Fig. \ref{chargetime-capacity-time}.

The transmitting energy is fixed, but the total discharging energy increases with the increasement of the receiver number during a charging process. Thus, the average SOC $R_{soc}$ decreases as the receiver number increases no matter how the charging duration changes in Fig. \ref{chargetime-capacity-time}. What's more, when the receiver number is same, the average SOC increases as prolonging the charging duration.

     \begin{figure}[!t]
	\centering
    \includegraphics[scale=0.6]{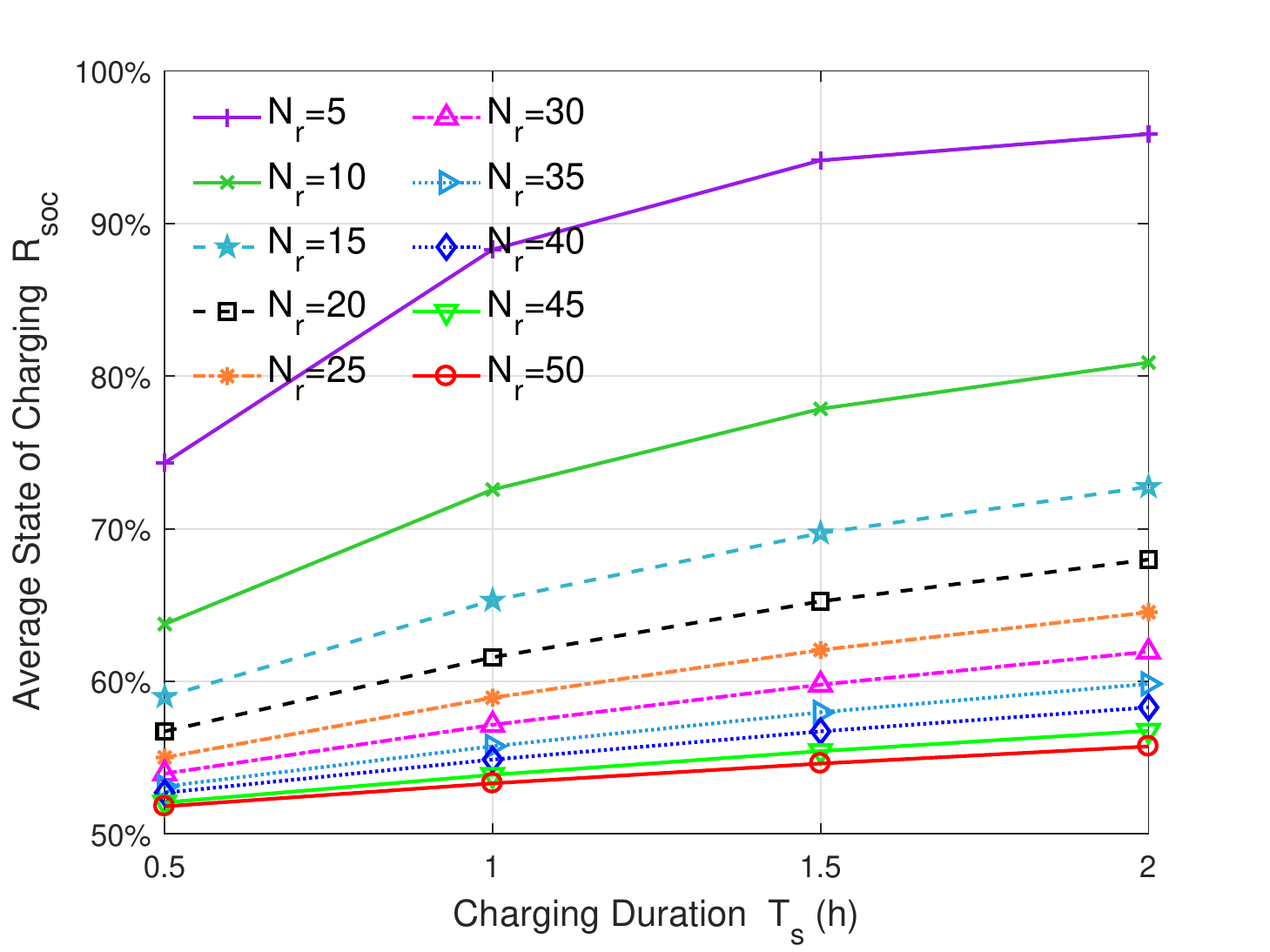}
	\caption{Average State of Charging vs. Charging Duration (Different Receiver Number)}
    \label{chargetime-capacity-number}
    \end{figure}

In addition, the changes of the receiver average SOC $R_{soc}$ over charging duration with different receiver number are depicted in Fig. \ref{chargetime-capacity-number}. From Fig. \ref{chargetime-capacity-number}, the average SOC increases with the extension of charging duration. The average SOC $R_{soc}$ decreases as the receiver number increases when the charging duration is same. Besides, when increasing the receiver number, the difference between total charging energy and total discharging energy reduces, so $R_{soc}$ grows slower over charging duration.

Based on the above analysis, the receiver's QoCS is better over the charging duration. However, when the receiver number increases, the QoCS is worse gradually. Thus, a reasonable strategy should be formulated to control the number of receivers charging simultaneously in the PtMP RBC system, and the charging duration should be extended if the charging conditions permit.

\begin{itemize}
  \item Earning Analysis
\end{itemize}

\begin{figure}[!t]
	\centering
    \includegraphics[scale=0.6]{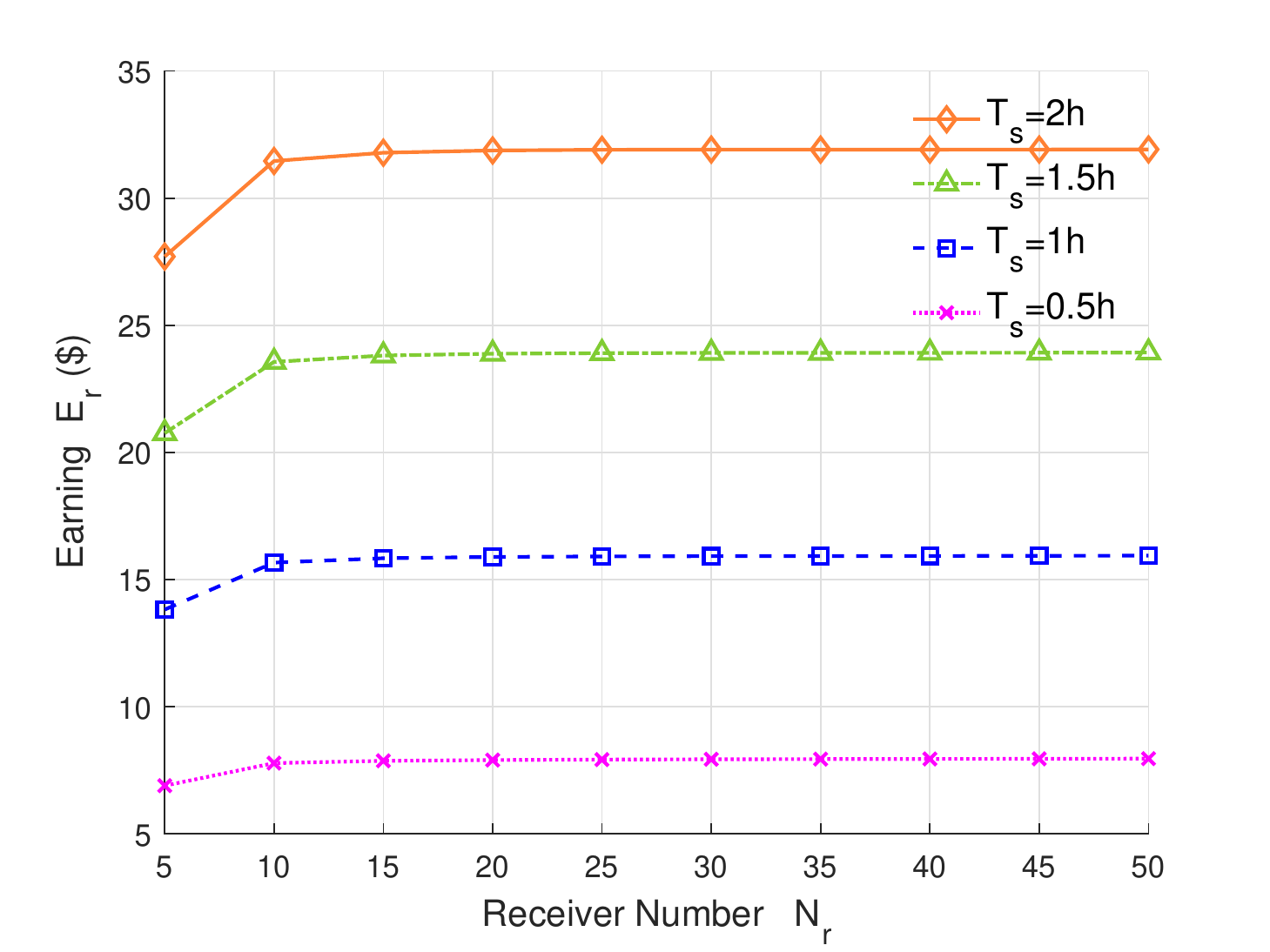}
	\caption{Earning vs. Receiver Number (Different Charging Duration)}
    \label{chargetime-capacity-timeearn}
    \end{figure}

The earning is determined by the CPS and the allocation of the transmitting power. The change trend of the earning with different receiver number under 0.5h, 1h, 1.5h, and 2h charging is depicted in Fig. \ref{chargetime-capacity-timeearn}.

In Fig. \ref{chargetime-capacity-timeearn}, the earning increases firstly, and then keeps the fixed and maximal value approximatively as the receiver number increases. The transmitting power is not fully allocated when the receiver number is small (e.g. 5, 10), while the allocated power is the maximal transmitting power when the receiver number is large (for example 40, 45). In addition, the earning gains as the charging duration prolongs, for example, the earning of charging 2h is greater than that of 1.5h.

\begin{figure}[!t]
	\centering
    \includegraphics[scale=0.6]{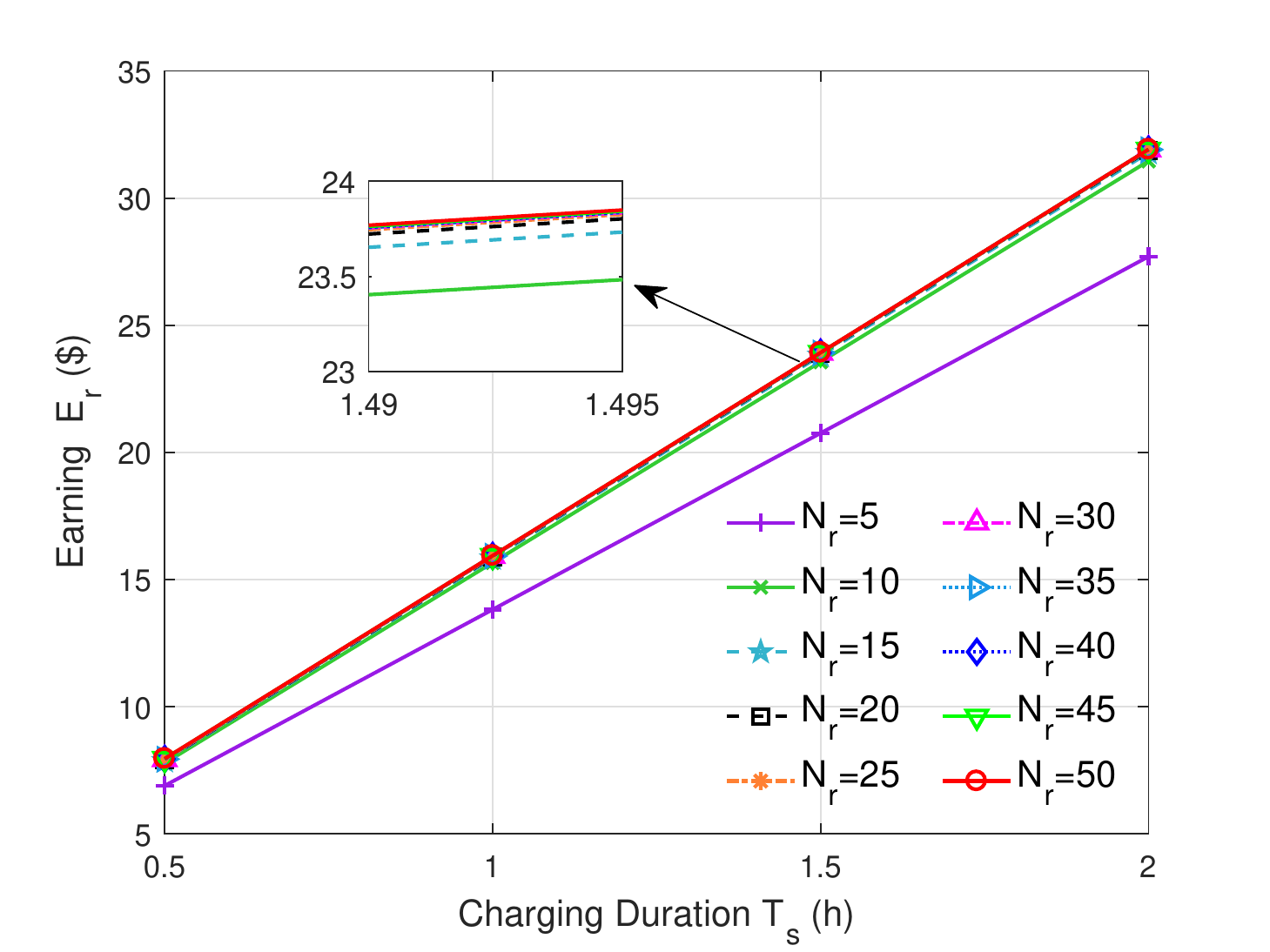}%scale=0.665
	\caption{Earning vs. Charging Duration (Different Receiver Number)}
    \label{chargetime-capacity-numearn}
    \end{figure}

The change trend of the earning with different receiver number under different charging duration is shown in Fig. \ref{chargetime-capacity-numearn}. From Fig. \ref{chargetime-capacity-numearn}, the earning increases rapidly with the extension of charging duration. What's more, when the receiver number is large, the total transmitting power is allocated fully, and the earning reaches the maximum approximately. Therefore, the growth trend of the earning is almost same and the value is similar when the receiver number is equal and greater than 10.

On this basis, the earning will increase when prolonging the charging duration and increasing the receiver number. However, increasing the receiver number does not significantly promote the increasement of earning when it is over a specific value. Thus, the methods to increase the earning include: a) prolong the charging duration, and b) limit the receiver number under a specific value.

Depending on the above analysis, we summarize the performance analysis results of the HPC scheduling algorithm as follows:

1) The charging performance of HPC algorithm is better than that of the RRC algorithm by comparing the QoCS and the earning. Thus, the HPC scheduling algorithm is more suitable for the PtMP RBC service, and its performance is superior.

2) The performance of the HPC scheduling charging algorithm is affected by the charging duration and the receiver number. The QoCS is better with the extension of the charging duration and the decreasement of the receiver number, while the earning increases as the charging duration prolongs and the receiver number increases.

3) The methods to improve the performance of the HPC scheduling algorithm include: a) Extend the charging duration appropriately if the charging conditions permit. b) Control the receiver number within a reasonable range, since the large receiver number is not conducive to the QoCS improvement, and the small receiver number goes against the earning increasement.
%%%%%%%%%%%%%%%%%%%%%%%%%%%%%%%%%%%%%%%%%%%%%%%%%%%%%%%%%%%%%%%%%%%%%%%%%%%%%%%%%%%%%%%%%%%%%%%%%%%%%%%%%%%%%%%%%%%%
\section{Conclusions}\label{Section5}

In the Point-to-Multipoint (PtMP) Resonant Beam Charging (RBC) system with the Charging Pricing Strategy (CPS), multiple receivers with various charging power, State of Charging (SOC), and using status can be charged simultaneously. We propose the High Priority Charge (HPC) scheduling algorithm to maximize the system earning with each receiver's Quality of Charging Service (QoCS) guarantee.

Based on the CPS and receiver status, we present the HPC scheduling algorithm to assign priorities for the receivers, and charge the receivers from high priority to low. To implement the HPC scheduling algorithm, we quantify the charging and discharging stages and relevant parameters, respectively. Then, we present the implementation pseudo code and steps on this basis. Relying on the simulation analysis, we find that the HPC algorithm can guarantee receiver's QoCS and achieve higher earning compared with the Round-Robin Charge (RRC) algorithm. In addition, to improve the PtMP RBC service, we figure out that the receiver number should be controlled within a reasonable range, and the charging duration should be extended as long as possible.

There are still some open issues worth further study, for example:

    \begin{itemize}
      \item The Charging Pricing Strategy (CPS) of the RBC system can be further formulated, including the pricing of the charging capacity and the detailed charging grade.
      \item The performance of the HPC scheduling algorithm considering different influencing factors, e.g., the transmitting power, the charging efficiency and so on, need to be studied.
      \item The factors that determine the receiver priority can be more diverse, for example, the quality of the charging link and the receiver's power consumption.
    \end{itemize}

%%%%%%%%%%%%%%%%%%%%%%%%%%%%%%%%%%%%%%%%%%%%%%%%%%%%%%%%%%%%%%%%%%%%%%%%%%%%%%%%%%%%%%%%%%%%%%%%%%%%%%%%%%%%%%%%%%%%
\section*{Acknowledgements}

To begin with, we would like to show our deepest gratitude to our friends and colleagues, Hao Deng and Mingliang Xiong, who have provided us with valuable suggestions in improving this paper. Next, our sincere appreciation also goes to the co-worker, Aozhou Wu, who has participated this study with polishing the figures in this paper.

\bibliographystyle{IEEEtran}
\bibliographystyle{unsrt}
\bibliography{references}

\begin{thebibliography}{10}
\providecommand{\url}[1]{#1}
\csname url@samestyle\endcsname
\providecommand{\newblock}{\relax}
\providecommand{\bibinfo}[2]{#2}
\providecommand{\BIBentrySTDinterwordspacing}{\spaceskip=0pt\relax}
\providecommand{\BIBentryALTinterwordstretchfactor}{4}
\providecommand{\BIBentryALTinterwordspacing}{\spaceskip=\fontdimen2\font plus
\BIBentryALTinterwordstretchfactor\fontdimen3\font minus
  \fontdimen4\font\relax}
\providecommand{\BIBforeignlanguage}[2]{{%
\expandafter\ifx\csname l@#1\endcsname\relax
\typeout{** WARNING: IEEEtran.bst: No hyphenation pattern has been}%
\typeout{** loaded for the language `#1'. Using the pattern for}%
\typeout{** the default language instead.}%
\else
\language=\csname l@#1\endcsname
\fi
#2}}
\providecommand{\BIBdecl}{\relax}
\BIBdecl

\bibitem{gubbi2013internet}
J.~Gubbi, R.~Buyya, S.~Marusic, and M.~Palaniswami, ``Internet of {T}hings
  ({I}o{T}): A vision, architectural elements, and future directions,''
  \emph{Future generation computer systems}, vol.~29, no.~7, pp. 1645--1660,
  Sept. 2013.

\bibitem{wu2014cognitive}
Q.~Wu, G.~Ding, Y.~Xu, S.~Feng, Z.~Du, J.~Wang, and K.~Long, ``Cognitive
  {I}nternet of {T}hings: {A} new paradigm beyond connection,'' \emph{IEEE
  Internet of Things Journal}, vol.~1, no.~2, pp. 129--143, Apr. 2014.

\bibitem{ding2018amateur}
G.~Ding, Q.~Wu, L.~Zhang, Y.~Lin, T.~A. Tsiftsis, and Y.-D. Yao, ``An amateur
  drone surveillance system based on the cognitive {I}nternet of {T}hings,''
  \emph{IEEE Communications Magazine}, vol.~56, no.~1, pp. 29--35, Jan. 2018.

\bibitem{wang2018power}
H.~Wang, G.~Ding, F.~Gao, J.~Chen, J.~Wang, and L.~Wang, ``Power control in
  {UAV}-supported ultra dense networks: Communications, caching, and energy
  transfer,'' \emph{IEEE Communications Magazine}, vol.~56, no.~6, pp. 28--34,
  Jun. 2018.

\bibitem{hui2014critical}
S.~Y.~R. Hui, W.~Zhong, and C.~K. Lee, ``A critical review of recent progress
  in mid-range wireless power transfer,'' \emph{IEEE Transactions on Power
  Electronics}, vol.~29, no.~9, pp. 4500--4511, Sept. 2014.

\bibitem{lu2016wireless}
X.~Lu, P.~Wang, D.~Niyato, D.~I. Kim, and Z.~Han, ``Wireless charging
  technologies: Fundamentals, standards, and network applications,'' \emph{IEEE
  Communications Surveys \& Tutorials}, vol.~18, no.~2, pp. 1413--1452, 2016.

\bibitem{wirelesstechniques}
X.~Lu, D.~Niyato, P.~Wang, D.~I. Kim, and Z.~Han, ``Wireless charger networking
  for mobile devices: Fundamentals, standards, and applications,'' \emph{IEEE
  Wireless Communications}, vol.~22, no.~2, pp. 126--135, Apr. 2015.

\bibitem{electromagnetic}
A.~Costanzo, M.~Dionigi, D.~Masotti, M.~Mongiardo, G.~Monti, L.~Tarricone, and
  R.~Sorrentino, ``Electromagnetic energy harvesting and wireless power
  transmission: A unified approach,'' \emph{Proceedings of the IEEE}, vol. 102,
  no.~11, pp. 1692--1711, Nov. 2014.

\bibitem{liu2016dlc}
Q.~Liu, J.~Wu, P.~Xia, S.~Zhao, W.~Chen, Y.~Yang, and L.~Hanzo, ``Charging
  unplugged: Will distributed laser charging for mobile wireless power transfer
  work?'' \emph{IEEE Vehcular Technology Magzine}, vol.~11, no.~4, pp. 36--45,
  Dec. 2016.

\bibitem{Qing2017}
Q.~Zhang, X.~Shi, Q.~Liu, J.~Wu, P.~Xia, and Y.~Liao, ``Adaptive distributed
  laser charging for efficient wireless power transfer,'' in \emph{Vehicular
  Technology Conference (VTC-Fall), 2017 IEEE 86th}, Sept. 2017, pp. 1--5.

\bibitem{feder2011mobility}
P.~M. Feder and K.~Livanos, ``Mobility aware policy and charging control in a
  wireless communication network,'' U.S. Patent 8\,086\,216, Dec., 2011.

\bibitem{foottit2011system}
T.~A. Foottit, Y.~Li, and M.~Jones, ``System and method for providing
  interoperability between diameter policy control and charging in a 3{GPP}
  network,'' U.S. Patent 7\,937\,300, May, 2011.

\bibitem{menasce2002qos}
D.~A. Menasc{\'e}, ``Qo{S} issues in web services,'' \emph{IEEE internet
  computing}, vol.~6, no.~6, pp. 72--75, Dec. 2002.

\bibitem{sood2017current}
K.~Sood, S.~Yu, and Y.~Xiang, ``Are current resources in {SDN} allocated to
  maximum performance and minimize costs and maintaining {Q}o{S} problems?'' in
  \emph{Proceedings of the Australasian Computer Science Week Multiconference},
  Feb. 2017, p.~42.

\bibitem{fang2018}
W.~Fang, Q.~Zhang, Q.~Liu, J.~Wu, and P.~Xia, ``Fair scheduling in resonant
  beam charging for iot devices (in press),'' \emph{IEEE Internet of Things
  Journal}, Jul. 2018, \mbox{doi}:\url{10.1109/JIOT.2018.2853546}.

\bibitem{Iphone8-review}
G.~Beavis, ``i{P}hone8 {P}lus review,'' [Onilne]. Available:
  \url{https://www.techradar.com/reviews/iphone-8-plus-review/4}, May 2018.

\bibitem{Iphone8-review2}
M.~Parker, ``i{P}hone8 {P}lus - battery life and verdict,'' [Onilne].
  Available:
  \url{http://www.trustedreviews.com/reviews/iphone-8-plus-battery-life-and-verdict},
  Jun. 2018.

\bibitem{Iphone8}
C.~Zibreg, ``\$74 to get i{P}hone fast charging is a waste of money—keep
  using your ipad chargers,'' [Onilne]. Available:
  \url{http://www.idownloadblog.com/2017/10/10/74-to-get-iphone-fast-charging-is-a-waste-of-money-keep-using-your-ipad-chargers/},
  Jun. 2018.

\bibitem{Iphone82}
C.~de~Looper and K.~Wiggers, ``Everything you need to know about charging your
  i{P}hone8 or i{P}hone8 {P}lus,'' [Onilne]. Available:
  \url{https://www.digitaltrends.com/mobile/apple-iphone-8-fast-wireless-charging/},
  Jun. 2018.

\bibitem{carroll2010analysis}
A.~Carroll and G.~Heiser, ``An analysis of power consumption in a smartphone,''
  in \emph{USENIX Annual Technical Conference}, vol.~14, 2010, p.~21.

\bibitem{murmuria2012mobile}
R.~Murmuria, J.~Medsger, A.~Stavrou, and J.~M. Voas, ``Mobile application and
  device power usage measurements,'' in \emph{2012 IEEE Sixth International
  Conference on Software Security and Reliability (SERE)}, Aug. 2012, pp.
  147--156.

\bibitem{zhang2018distributed2}
Q.~Zhang, W.~Fang, Q.~Liu, J.~Wu, P.~Xia, and L.~Yang, ``Distributed laser
  charging: A wireless power transfer approach (in press),'' \emph{IEEE
  Internet of Things Journal}, Jun. 2018,
  \mbox{doi}:\url{10.1109/JIOT.2018.2851070}.

\end{thebibliography}

\end{document}